\documentclass[preprint,showpacs,showkeys,amsmath,amssymb,floatfix,nofootinbib,longbib]{revtex4-1}
\usepackage{graphicx}
\usepackage{bbm}
\usepackage{bm}
\usepackage{color}
\usepackage[pdftex,colorlinks]{hyperref}

\def\P{\ensuremath{\mathcal{P}}}
\def\id{\ensuremath{\mathbbm{1}}}
\DeclareMathOperator{\Tr}{Tr}
\begin{document}
\title{Measurement of a qubit and measurement with a qubit}
\author{Antonio \surname{Di Lorenzo}}
\email{dilorenzo@infis.ufu.br}
\affiliation{Instituto de F\'{\i}sica, Universidade Federal de Uberl\^{a}ndia,\\
 38400-902 Uberl\^{a}ndia, Minas Gerais, Brazil}
\begin{abstract}
Generally, the measurement process consists in coupling a system to a detector that can give a continuous output. 
However, it may be interesting to use as a detector a system with a discrete spectrum, especially in view of applications 
to quantum information. 
Here, we study 1) a two-level system measuring another two-level system (qubit); 
2) a generic system measuring a qubit; 
3) a qubit measuring a generic system.
The results include the case when a postselection on the measured system is made. 
We provide the exact solution, and also a controlled expansion in the coupling parameter, giving formulas valid in the weak measurement regime for arbitrary preparation and postselection. 
The concept of generalized Wigner functions is introduced. 
\end{abstract}
\maketitle
\section{Introduction.}
Two-level systems are important because on one hand they realize the simplest non-trivial quantum systems, 
and on the other hand because they can be used as qubits, the basic units of quantum computation. 
While studies on quantum computation have relied mostly on projective measurements, it is interesting to 
consider more general measurements. 
The most general measurements are the positive-operator valued measures \cite{Davies1976,Helstrom1976,Holevo1982,Kraus1983,Ludwig1983,Busch1995}, 
that reduce, in particular, to postselected weak measurements, introduced in Ref.~\cite{Aharonov1988}, 
for a weak enough coupling between system and probe, if the former is subsequently measured projectively. 

Weak measurements are not just a theoretical tool, 
but they may be useful: for instance, it has been shown how weak measurements allow the reconstruction 
of the wave function \cite{Lundeen2011} or of the Bohmian trajectories \cite{Kocsis2011}

Weak measurements rely on a perturbative expansion of the propagator, and the postselection implies that, 
according to Bayes' rule, the joint probability of output and postselection be divided by the marginal 
probability of successful postselection. While this latter probability can be expanded in a Taylor series, its inverse 
may not, when the preparation and the postselection are nearly orthogonal, making the lower order term zero. 
Hence, one should be more careful in making a perturbative expansion in this case \cite{Wu2011,DiLorenzo2012a,DiLorenzo2012j,Kofman2012}. 
This expansion is sometimes called improperly ``non-perturbative'' \cite{Kofman2012}, but it actually breaks down for larger values of the coupling parameter. A more appropriate term would be perhaps ``non-polynomial expansion'', as the conditional 
probability can be written as a rational function in the coupling \cite{DiLorenzo2012a}. 

Two-level systems, however, are simple enough to allow for an exact treatment, hence there is no need to confine 
ourselves either to the strong or to the weak measurement regime. Indeed, sequential measurements of noncommuting 
variables have been shown to allow quantum state tomography,  if one stays in the intermediate 
coupling regime \cite{DiLorenzo2013a}. 
In the following we shall consider measurements followed by a postselection, but treat the problems exactly, 
while we shall provide approximate formulas valid in the weak measurement regime for ease of comparison with 
the previous results. 

Finally, we shall address two-level systems as ``qubits'', but we shall use the language of spin 1/2, 
not the language of quantum computation, with an exception in Section \ref{sec:qq}. The reason for doing so is that 
quantum computation relies on a special basis, the so-called computational basis, which conceals the invariance of the 
formulas. Spin 1/2 systems, on the other hand, allow to express significant quantities through geometrical expressions that 
are manifestly invariant, as they rely on the scalar products of vectors. 

The main results of this paper are:
\begin{enumerate}
\item
Equation \eqref{eq:qqcond}, which gives the statistics of a qubit when measured by another qubit.
\item
Equation \eqref{eq:avfo}, which gives the statistics of a qubit when measured by a general system.
\item
Equation \eqref{eq:exsolqg}, which gives the statistics of a general system when measured by a qubit.
\end{enumerate}
All these results are exact, apply for any coupling strength, for any preparation of the system and of the probe, 
for any postselection of the system, and 
for any readout basis of the probe. 
%
\section{General description of a measurement}
A measurement proceeds in the following way: 
(i) prepare the detecting system, which we call the probe for brevity,  in the ready state $\rho_\mathrm{P}$, uncorrelated to the initial state of the system to be measured $\rho_\mathrm{i}$, so that 
the total state is $\rho_\mathrm{P}\otimes\rho_\mathrm{i}$; (ii) have the measured system and the probe interact through a 
Hamiltonian $H_\mathrm{int}$; (iii) read the output of the probe in a properly chosen basis $\Hat{E}(O)$, where $\Hat{E}(O)$ is a family of nonnegative operators satisfying $\int d\mu(O) \Hat{E}(O)=1$, with $\mu$ a Lebesgues-Stieltjes measure accounting for a continuous, discrete, or mixed output spectrum \cite{DiLorenzo2012k}. 
The case most commonly studied is the one where $\Hat{E}(O)=|Q=O\rangle\langle Q=O|$, i.e. the readout of the detector is ideal, it is an 
eigenstate of a suitably chosen operator $\Hat{Q}$. In general, however, $\Hat{E}(O)$ are proportional to mixed state, $\Hat{E}(O)=\mathrm{pr}(O) \Hat{\rho}(O)$, 
with $\mathrm{pr}(O)$ being the prior probability of observing an outcome $O$. Often, the prior probability is not normalized, $\int d\mu(O) \mathrm{pr}(O)=\infty$, but 
this is not a problem: what matters are the ratios $\mathrm{pr}(O)/\mathrm{pr}(O')$ that give the relative probability of observing an output $O$ rather than an output $O'$, a priori, i.e. when no measurement is being made. For instance, a common case is that all output are equiprobable a priori, so that $\mathrm{pr}(O)=1$. If instead 
the probe has a different resolution $r_n$ for different outputs $O_n$, $\mathrm{pr}(O_n)\propto r_n$. 
For ease of comparison with the results of the previous literature, we shall consider the case of an ideal output, 
$\Hat{E}(O)=|Q=O\rangle\langle Q=O|$. 

In a strong measurement $H_\mathrm{int}$, $\rho_\mathrm{P}$, and $\Hat{E}(O)$ are chosen in such a way that, if initially the system is in the eigenstate $|A\rangle$ of $\Hat{A}$, 
the detector ends up in $\rho(O=f(A))$, with $f(A)$ a given function of $A$ establishing a correspondence between the output and the state of the measured system, 
and in particular $f(A)=\lambda A$ for a linear measurement.
This strong measurement is obtained under the following conditions
\begin{enumerate}
\item  Choose a bilinear interaction $H_\mathrm{int}=\lambda \delta(t) \Hat{A} \Hat{P}$, where $\Hat{A}$ is the observable of the system we wish to measure and $\Hat{P}$ is an observable of the probe. 
\item Prepare the probe in a state $|Q=0\rangle$ such that $|Q_A\rangle\equiv \exp(i\lambda A\Hat{P})|Q=0\rangle$ are orthogonal for different eigenvalues $A$ of $\Hat{A}$.  
\item Reading the probe in a basis containing $|Q_A\rangle$. 
\end{enumerate}
Notice that $\Hat{Q}$ and $\Hat{P}$ are not being assumed to be position and momentum of a pointer, nor to  be canonically conjugated variables, as in the following we shall treat the case of a finite-dimensional Hilbert space.  

In general, the one-to-one correspondence between output of the probe and state of the system does not hold, either because of intrinsic limitations in the preparation 
and readout operation, or because of intentional deviations from the ideal case with the goal of reducing the measurement back-action at the expense of the reliability. 
In the following, we shall drop hypotheses 2 and 3.

\begin{table}[h]
\begin{tabular}{|c|l|c|}
\hline
Symbol&Meaning&Restrictions\\
\hline
$\Hat{P}$&Write-in variable of the probe&None\\
\hline
$\Hat{Q}$&Readout variable of the probe&None\\
\hline
$\rho_\mathrm{P}$&Initial state of the probe&Non-negative, trace one operator\\
\hline
$\rho_\mathrm{P|f}$&Final state of the probe, given a postselection in $E_\mathrm{f}$&Non-negative, trace one operator\\
\hline
$\Hat{\boldsymbol{\tau}}$&Spin operator of a qubit probe&None\\
\hline
$\mathbf{n}$&Initial polarization of a qubit probe&$|\mathbf{n}|\le 1$, $\rho_\mathrm{P}=(1+\mathbf{n}\cdot\Hat{\boldsymbol{\tau}})/2$\\
\hline
$\mathbf{p}$&Orientation of $\Hat{P}$ for a qubit probe&$|\mathbf{p}|= 1$, $\Hat{P}=(1-\mathbf{p}\cdot\Hat{\boldsymbol{\tau}})/2$\\
\hline
$\mathbf{q}$&Orientation of the readout of a qubit probe&$|\mathbf{q}|= 1$\\
\hline
$\Hat{A}$&Variable of the system to be measured&None\\
\hline
$\rho_\mathrm{i}$&Preparation state of the system&Non-negative, trace one operator\\
\hline
$E_\mathrm{f}$&Postselected state of the system&Positive operator, $E_\mathrm{f}\le 1$\\
\hline
$\rho_\mathrm{f}$&Normalized postselected state of the system&$\rho_\mathrm{f} =E_\mathrm{f}/\Tr{(E_\mathrm{f})}$\\
\hline
$\Hat{\boldsymbol{\sigma}}$&Spin operator of a qubit system&None\\
\hline
$\mathbf{m}_\mathrm{i}$&Preparation polarization of a qubit system&$|\mathbf{m}_\mathrm{i}|\le 1$, $\rho_\mathrm{i}=(1+\mathbf{m}_\mathrm{i}\cdot\Hat{\boldsymbol{\sigma}})/2$\\
\hline
$\mathbf{m}_\mathrm{f}$&Postselection polarization  of a qubit system&$|\mathbf{m}_\mathrm{f}|\le 1$, $\rho_\mathrm{f}=(1+\mathbf{m}_\mathrm{f}\cdot\Hat{\boldsymbol{\sigma}})/2$\\
\hline
$\mathbf{a}$&Polarization of $\Hat{A}$ for a qubit &$|\mathbf{a}|= 1$, $\Hat{A}=\mathbf{a}\cdot\Hat{\boldsymbol{\sigma}}$\\
\hline
\end{tabular}
\caption{\label{table} The symbols used throughout the paper.}
\end{table}

Furthermore, a second measurement on the system can be realized subsequently, allowing to select the data 
of the first measurement depending on the outcome of the second one. This procedure is known as postselection. 
The system can be postselected in a pure or mixed state. The postselected state $E_\mathrm{f}$ is a positive operator that satisfies $E_\mathrm{f}\le 1$ and that may be not normalized,  but this has no consequences as far as the conditional probability is considered 
$Prob(O|postselection)=Prob(O \& postselection)/Prob(postselection)$. In particular, the case when no postselection is made is obtained for $E_\mathrm{f}=\id$. 
The postselection in a mixed state can be achieved either by making the second measurement a Positive Operator Valued measurement \cite{Wiseman2002}, or by making the postselection a probabilistic process \cite{DiLorenzo2012a,DiLorenzo2012e}. 
Notice that if we multiply $E_\mathrm{f}$ by a positive constant $c$, keeping $c E_\mathrm{f}\le 1$ the normalized postselected state does not change.  
The probability of a successful postselection, however, is proportional to $\Tr{E_\mathrm{f}}$. Hence, an optimal choice is to choose the maximal $c$, the one that makes the maximum eigenvalue of $E_\mathrm{f}$ equal 1. For instance, let us consider the case where no postselection is done, yet a fraction $x$ of the data is discarded by mistake or due to a malfunction of the apparatus. Then, ${E_\mathrm{f}}=(1-x)\id$, the normalized postselected state is $\rho_\mathrm{f}=\id/N$ with $N$ the dimension of the Hilbert space, the same that would be obtained for $x=0$, but in this case all of the data would be used.  

%
\section{A qubit measuring a qubit.}\label{sec:qq}

For the sake of exposition, let us take momentarily 
the $Z$-axis along the direction of the spin component being measured, i.e. $\Hat{A}=(1-\Hat{\sigma}_z)/2$. 
The shift and the rescaling are done for convenience, and are not essential for the results to follow. 
We choose the readout basis of the detector to be 
along the $Z$-axis as well.\footnote{Notice that the two $Z$-axes are not necessarily the same. Another way to state this is that we consider the eigenstates of the measured quantity $\Hat{A}$, fix arbitrarily the relative phase, and label them 
$|+\rangle,|-\rangle$; the same is done with the eigenstates of the readout operator $\Hat{Q}$, which are labelled as 
$|\tau_z=+\rangle, |\tau_z=-\rangle$.}  
Then, for an ideal measurement, the operator $\Hat{P}$ must be a spin component orthogonal to $\tau_z$ (we indicate with $\sigma_j$ the operators on the system, and $\tau_j$ the operators on the detector). Without loss of generality, we can put $\Hat{P}=(1-\Hat{\tau}_x)/2$. 
It is readily verified that if the coupling constant is $\lambda=\pi$ and the ready state is $|\tau_z=+1\rangle$, then the 
measurement is ideal. 
Indeed the time-evolution operator is 
\begin{equation}
\mathcal{U}=\exp{\left[i\frac{\pi}{4}(1-\Hat{\sigma}_z)(1-\Hat{\tau}_x)\right]}.
\label{eq:evol}
\end{equation}
When the system is in the state $|\Hat{\sigma}_z=+\rangle$, 
$\langle \Hat{\sigma}_z=+|\mathcal{U}|\Hat{\sigma}_z=+\rangle$ is the identity over the 
detector; when the system is in the state $|\Hat{\sigma}_z=-\rangle$, 
$\langle \Hat{\sigma}_z=-|\mathcal{U}|\Hat{\sigma}_z=-\rangle=\Hat{\tau}_x$, 
so that, if the detector is initially in the pure state $\rho_\mathrm{P}=|R\rangle\langle R|$, 
$|R\rangle=|\tau_z=+\rangle$ it ends up in $|\tau_z=-\rangle$ 
\footnote{One can choose as well the opposite ready state $|\tau_z=-\ensuremath{\rangle}$.}.
In quantum information, the operator $\mathcal{U}$ in Eq.~\eqref{eq:evol} is known as a controlled NOT: 
if we identify $\tau_z=+$ with the logic symbol $0$ and $\tau_z=-$ with $1$, then $\mathcal{U}$ takes 0 to 1 and 
vice versa for $\Hat{\sigma}_z=-$ (i.e., the control bit is in the ``true'' state 1), while it does nothing for $\Hat{\sigma}_z=+$. 
\begin{figure}
\includegraphics[width=7in]{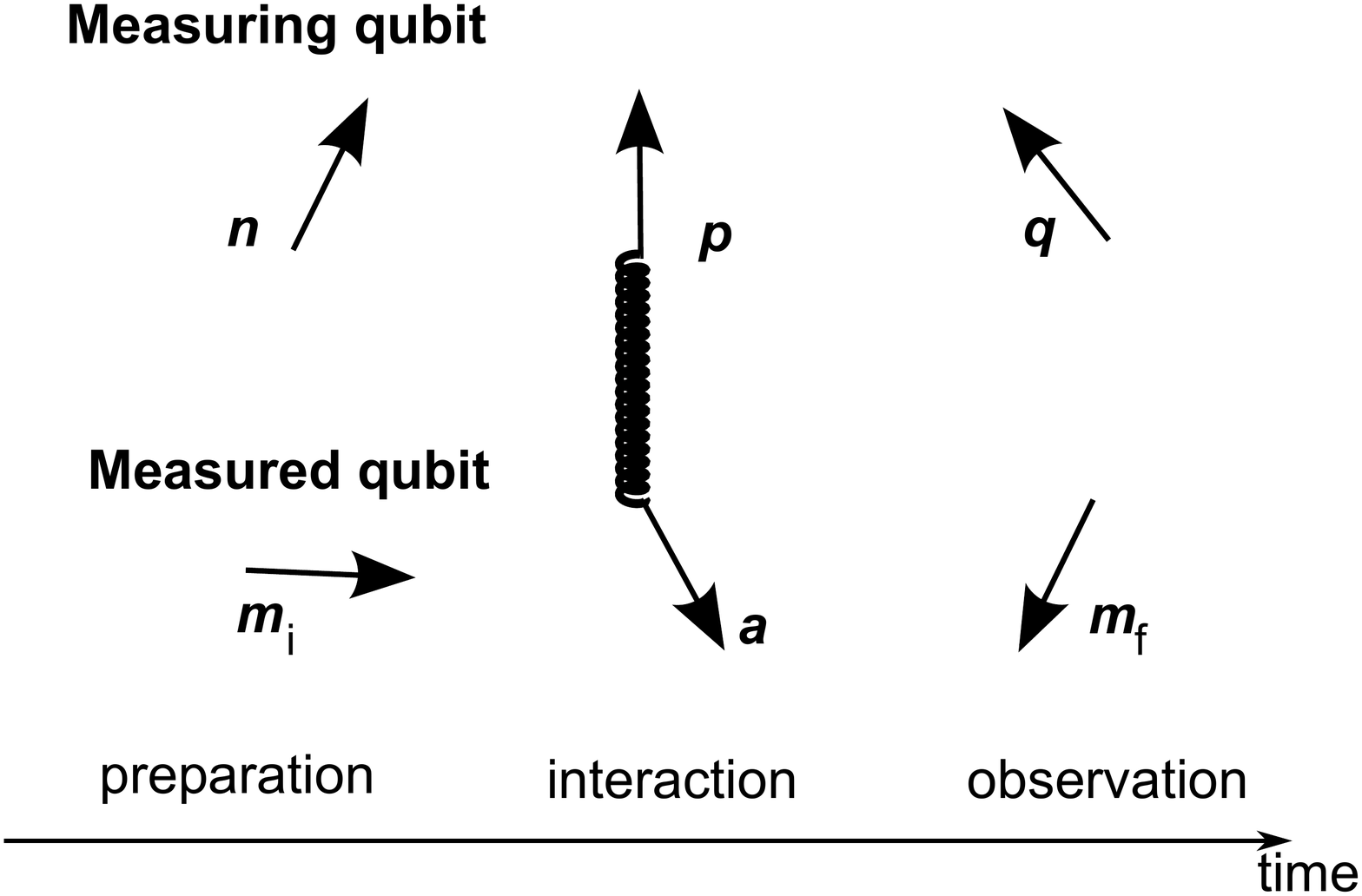}
\caption{Scheme of the measurement of a spin 1/2 using another spin 1/2.}
\end{figure}
\subsection{Fixed coupling, arbitrary probe preparation}
Along the lines of Ref.~\cite{Pryde2005}, we consider a weak measurement, not in the sense that $\lambda\ll 1$, 
but in the sense that the ready state of the detector is left arbitrary, with $\lambda$ fixed to $\pi$. 
Hence, the detector is considered to be initially prepared in a general state 
$\rho_\mathrm{P}=(1+\mathbf{n}\cdot\Hat{\boldsymbol{\tau}})/2$ with $|\mathbf{n}|\le 1$. 
Averages over the initial state will be denoted by a bar. 
Since the only non-trivial averages for a spin-1/2 are the spin components, we shall introduce the average vector 
$\overline{\Hat{\boldsymbol{\tau}}}=\mathbf{n}$. 
Furthermore, to keep equations manifestly invariant, we replace the $X$-direction of the probe qubit 
with a generic unit vector $\mathbf{p}$, and we write $\Hat{A}=\mathbf{a}\cdot\Hat{\boldsymbol{\sigma}}$ 
with $\mathbf{a}$ a unit vector 
[notice that the quantity appearing in the interaction Hamiltonian is $(1-\Hat{A})/2$, which results 
in a trivial systematic shift and rescaling of the readout]. 

In Ref.~\cite{DiLorenzo2012a} we have proved that the statistics of the readout can be written in terms 
of the normal weak values 
\begin{equation}
\alpha_{m,n}=
\mathrm{Tr}(\Hat{A}^m E_\mathrm{f} \Hat{A}^n\rho_\mathrm{i}), 
\label{eq:normwv}
\end{equation} 
where $\Hat{A}$ is the observable being measured ($\Hat{\sigma}_z$ in our case). 
For a spin 1/2, there are only three parameters: the positive real numbers $\alpha_{0,0}\equiv\omega$ and 
$\alpha_{1,1}\equiv\beta$, and the complex 
quantity $\alpha_{0,1}\equiv\alpha$.  These quantities are already written in invariant form. Their geometric expressions are 
\begin{align}
\omega=&\ \frac{1}{2}\Tr{(E_\mathrm{f})}
\left(1+\mathbf{m}_\mathrm{i}\!\cdot\!\mathbf{m}_\mathrm{f}\right),\\
\alpha=&\ \frac{1}{2}\Tr{(E_\mathrm{f})}{\left(\mathbf{m}_\mathrm{i}+\mathbf{m}_\mathrm{f}+i\mathbf{m}_\mathrm{i}\times\mathbf{m}_\mathrm{f}\right)\cdot\mathbf{a}},\\
\beta =&\ \frac{1}{2}\Tr{(E_\mathrm{f})}\left({1-\mathbf{m}_\mathrm{i}\!\cdot\!\mathbf{m}_\mathrm{f}+2 \mathbf{m}_\mathrm{f}\!\cdot\!\mathbf{a}\, \mathbf{a}\!\cdot\!\mathbf{m}_\mathrm{i}}\right) ,
\end{align}
where 
the preparation of the system and the (normalized) postselection are $\rho_\mathrm{i}=(1+\mathbf{m}_\mathrm{i}\cdot\Hat{\boldsymbol{\sigma}})/2$ and 
$\rho_\mathrm{f}=(1+\mathbf{m}_\mathrm{f}\cdot\Hat{\boldsymbol{\sigma}})/2$, with $|\mathbf{m}_\mathrm{i}|\le 1, |\mathbf{m}_\mathrm{f}|\le 1$. 

We consider the readout to be in an arbitrary basis $|\mathbf{q}:\tau\rangle$. 
The eigenstates of $\mathbf{a}\cdot\Hat{\boldsymbol{\sigma}}$ are denoted by $|\pm\rangle$. 
The joint probability of observing an output $\tau$ and making a successful postselection in $E_\mathrm{f}$ is 
\begin{align}
\P(E_\mathrm{f},\tau)&=
\langle+|E_\mathrm{f}|+\rangle\langle+|\rho_\mathrm{i}|+\rangle \langle \mathbf{q}:\tau|\rho_\mathrm{P}|\mathbf{q}:\tau\rangle
+\langle-|E_\mathrm{f}|-\rangle\langle-|\rho_\mathrm{i}|-\rangle 
\langle \mathbf{q}:\tau|\mathbf{p}\!\cdot\!\Hat{\boldsymbol{\tau}}\rho_\mathrm{P}\mathbf{p}\!\cdot\!\Hat{\boldsymbol{\tau}}|\mathbf{q}:\tau\rangle
\nonumber
\\
&+\langle+|E_\mathrm{f}|-\rangle\langle-|\rho_\mathrm{i}|+\rangle \langle \mathbf{q}:\tau|\mathbf{p}\!\cdot\!\Hat{\boldsymbol{\tau}}\rho_\mathrm{P}|\mathbf{q}:\tau\rangle
+\langle-|E_\mathrm{f}|+\rangle\langle+|\rho_\mathrm{i}|-\rangle \langle \mathbf{q}:\tau|\rho_\mathrm{P}\mathbf{p}\!\cdot\!\Hat{\boldsymbol{\tau}}|\mathbf{q}:\tau\rangle .
\label{eq:prob0}
\end{align}
The first line can be interpreted ``classically'', i.e., as if the measured value 
is a prepossessed property of the system: indeed, if the system has the value $+$, and this happens with probability 
$\langle+|\rho_\mathrm{i}|+\rangle$, it will be postselected in $E_\mathrm{f}$ with probability $\langle+|E_\mathrm{f}|+\rangle$, and since in this case 
the evolution of the detector is trivial, the probability of observing the output $\tau$ is 
$\langle\mathbf{q}:\tau|\rho_\mathrm{P}|\mathbf{q}:\tau\rangle$; if, instead, the system has initially the value 1, then the detector is rotated 
by an angle $\pi$ around the $X$-axis, yielding a probability $\langle \mathbf{q}:\tau|\Hat{\tau}_x\rho_\mathrm{P}\Hat{\tau}_x|\mathbf{q}:\tau\rangle$. 
The second line represents interference effects, does not allow a simple interpretation as the one above, and is at the origin of the strange behavior of the weak value. 

The four numbers $\langle\sigma|E_\mathrm{f}|\sigma'\rangle\langle\sigma'|\rho_\mathrm{i}|\sigma\rangle$ can be expressed in terms of 
the four real invariant parameters $\omega,\beta,\mathrm{Re}(\alpha)\equiv\alpha',
\mathrm{Im}(\alpha)\equiv\alpha''$, 
\begin{subequations}
\begin{align}
 \langle+|E_\mathrm{f}|+\rangle\langle+|\rho_\mathrm{i}|+\rangle =& \frac{1}{4}\left[\omega+\beta+2\alpha'\right],\\
 \langle-|E_\mathrm{f}|-\rangle\langle-|\rho_\mathrm{i}|-\rangle =& \frac{1}{4}\left[\omega+\beta-2\alpha'\right],\\
 \langle+|E_\mathrm{f}|-\rangle\langle-|\rho_\mathrm{i}|+\rangle =& \frac{1}{4}\left[\omega-\beta-2i\alpha''\right],\\
 \langle-|E_\mathrm{f}|+\rangle\langle+|\rho_\mathrm{i}|-\rangle =& \frac{1}{4}\left[\omega-\beta+2i\alpha''\right].
\end{align}
\label{eq:comps}
\end{subequations}

The probability of a successful postselection is found by summing Eq.~\eqref{eq:prob0} over $\tau$,\footnote{Quantum mechanics is a local theory, hence, since the detector and the probe do not interact any longer after 
$t=0$, the probability of postselection does not depend on what observable of the detector is measured, or on whether 
an observable is measured at all. This is a natural consequence of the invariance of the trace and of the factorization of the time-evolution operator for noninteracting systems.}
\begin{equation}\label{eq:ppost}
\P(E_\mathrm{f}) = \frac{1}{2}\left[(1+\mathbf{p}\!\cdot\!\mathbf{n})\omega+
(1-\mathbf{p}\!\cdot\!\mathbf{n})\beta\right] . 
\end{equation}

Next, we consider the conditional probability, obtained by dividing the joint probability of Eq.~\eqref{eq:prob0} by 
$\P(E_\mathrm{f})$,  
\begin{align}
\label{eq:genweakprob}
\mathcal{Q}(\tau):= 
 \frac{\P(E_\mathrm{f},\tau)}{\P(E_\mathrm{f})}.
\end{align}
Since both $\P(E_\mathrm{f},\tau)$ and $\P(E_\mathrm{f})$ are homogeneous functions of degree one in $\alpha_{m,n}$, 
we may divide both by $\omega$, eliminating one parameter. 
Accordingly, we define the canonical complex weak value $A_w=\alpha/\omega$ and the 
generalized (real) weak value $B_w=\beta/\omega$. For pure preparation and postselection 
$B_w=|A_w|^2$, otherwise\footnote{More precisely~\cite{DiLorenzo2012e}, $B_w=|A_w|^2$ whenever 
the eigenvectors of $\rho_\mathrm{f}$ with nonzero eigenvalue, $|f:k\rangle$,  and those of $\rho_\mathrm{i}$, $|i:k\rangle$, satisfy 
$\langle f:k|\Hat{A}|i:k'\rangle/\langle f:k|i:k'\rangle=constant, \forall k,k'$.} $B_w>|A_w|^2$. For brevity, $\mathrm{Re}(A_w)=A'_w$  and 
$\mathrm{Im}(A_w)=A''_w$. We note that when no postselection is made, $A_w=\langle \Hat{A}\rangle_\mathrm{i}$ and 
$B_w=\langle \Hat{A^2}\rangle_\mathrm{i}=1$.

As the outcome is binary, the average value of $\tau$ gives all the necessary information about $\mathcal{Q}$.
 Multiplying Eq.~\eqref{eq:prob0} 
by $\tau$, summing, substituting Eqs.~\eqref{eq:comps}, and dividing by Eq.~\eqref{eq:ppost}, we find 
\begin{align}
\langle\tau\rangle=& \frac{  
\mathbf{q}\!\cdot\!\mathbf{p}\,(1+ \mathbf{p}\!\cdot\!\mathbf{n})
+2A'_w\left(\mathbf{q}\!\cdot\!\mathbf{n}-
\mathbf{q}\!\cdot\!\mathbf{p}\, \mathbf{p}\!\cdot\!\mathbf{n}\right) 
-2A''_w(\mathbf{q}\!\times\!\mathbf{n})\!\cdot\!\mathbf{p}
-B_w 
\mathbf{q}\!\cdot\!\mathbf{p}\,(1- \mathbf{p}\!\cdot\!\mathbf{n})}
{(1+\mathbf{p}\!\cdot\!\mathbf{n})+B_w(1-\mathbf{p}\!\cdot\!\mathbf{n})} .
\end{align}
This is a general invariant formula that resumes the statistics of the measurement of a qubit with a qubit for fixed coupling $\lambda=\pi$. 
 
%

\subsubsection*{Applications}
In particular, let us consider a readout made in the basis $\mathbf{q}=\Hat{\mathbf{z}}$, 
and an interaction along $\Hat{\mathbf{x}}$, as in the first example and corresponding to the 
coordinate system chosen in Ref.~\cite{Pryde2005}. 
Then 
\begin{align}
\langle\tau\rangle=& \frac{ 
2A'_w\overline{{\tau_z}}+
2A''_w\overline{{\tau_y}}
}
{(1+\overline{{\Hat{\tau}_x}})+(1-\overline{{\Hat{\tau}_x}})B_w } .
\end{align}
In a strong measurement $\overline{\tau_z}=1$ , $\overline{\Hat{\tau}_x}=\overline{\tau_y}=0$. In a weak incoherent measurement, 
the detector is prepared in a mixture of eigenstates of $\tau_z$, yielding 
$\overline{\tau_z}<1$ , $\overline{\Hat{\tau}_x}=\overline{\tau_y}=0$, so that 
$\langle\tau\rangle=2A'_w \overline{\tau_z}/(1+B_w)$. In a sense, $\overline{\tau_z}$ represents the amplification factor. 
Following this consideration, Ref.~\cite{Pryde2005} divides the output by  $\overline{\tau_z}=2\gamma^2-1$ and obtains 
an ``amplification'', even though a two-level system is being used. Indeed, when the initial state of the detector is 
very close to the state $|\Hat{\tau}_x=+\rangle$, say in the $XZ$ plane and making an angle $\varepsilon$ with the $X$-axis 
($\sin{\varepsilon}=2\gamma^2-1$ in the notation of Ref.~\cite{Pryde2005}), 
\begin{align}
\langle\tau\rangle=& \frac{ 
2A'_w\sin{\varepsilon}
}
{(1+\cos{\varepsilon})+(1-\cos{\varepsilon})B_w } .
\end{align}
In the limit $\varepsilon\to 0$, $\langle\tau\rangle/\sin{\varepsilon}\to A'_w$. 
When the preparation and the postselection are nearly orthogonal, this limit can be very large. 
However, for finite $\varepsilon$, the approximate equality $\langle\tau\rangle/\sin{\varepsilon}\simeq  A'_w$ 
holds only as far as $\varepsilon^2 B_w\ll 1$. For nearly orthogonal preparation and postselection $B_w$ 
becomes very large (much larger than $A_w$, recall that $B_w\ge |A_w|^2$) and the second-order correction in the 
denominator can not be neglected. In other words, care must be taken to apply the limits $\varepsilon\to 0$ and 
$\Tr\{\rho_\mathrm{f}\rho_\mathrm{i}\}=\omega\to 0$ in the correct order.

Interestingly, for $\varepsilon\to \pi$, $\langle\tau\rangle/\sin{\varepsilon}\to A'_w/B_w$, so that preparing the detector 
in the other coherent state $|\Hat{\tau}_x=-1\rangle$ one can retrieve the value of $B_w$. 
Furthermore, the imaginary part of $A_w$ can be inferred by preparing the detector in a state with $\overline{\tau}_y=1$, 
$\overline{\tau}_z=\overline{\tau}_x=0$. 

A possible application is the following: 
Suppose we want to establish whether the initial state $\rho_\mathrm{i}$ is pure or mixed. 
We make a postselection in a pure state $\rho_\mathrm{f}$. By changing the preparation of the detector, we 
retrieve $A'_w$, $A''_w$, and $B_w$. If $B_w>(A'_w)^2+(A''_w)^2$, we conclude that the initial state of the system is mixed.  
\subsection{Arbitrary coupling and probe preparation}
Let us consider now a fixed preparation and a varying interaction strength. 
We shall momentarily choose a coordinate system where $\mathbf{a}=\mathbf{\Hat{z}}$, $\mathbf{p}=\mathbf{\Hat{x}}$, 
and $\mathbf{q}=\mathbf{\Hat{z}}$. 
We take the initial state of the detector to be, in the basis of eigenstates of $\Hat{\tau}_z$,  
\begin{equation}
\rho_\mathrm{P} = \frac{1}{2}\left(1+\mathbf{n}\cdot\Hat{\boldsymbol{\tau}}\right)=
\begin{pmatrix}w&\kappa e^{-i\delta}
\\
\kappa e^{i\delta}&1-w
\end{pmatrix} ,
\end{equation}
with $0\le w\le 1$, the probability of the probe being initially in the ``up'' state, 
$0\le \kappa\le \sqrt{w(1-w)}$ the coherence, and $\delta$ a phase. 
Thus $\mathbf{n}=2\kappa [\cos{(\delta)} \Hat{\mathbf{x}}+\sin{(\delta)}\Hat{\mathbf{y}}]+(2w-1)\Hat{\mathbf{z}}$. 
The interaction is 
\begin{equation}
H = -\hbar \delta(t) \frac{\lambda}{4} (1-\Hat{\sigma}_z) (1-\Hat{\tau}_x)
\end{equation}
If the detector has a continuous output, the regime of coherent weak measurement is obtained for 
$\lambda\ll \kappa$. Hence, one can either increase the coherence $\kappa$ (and with it the spread $\Delta=2\sqrt{w(1-w)}\ge 2\kappa$, because of the uncertainty relation) or decrease $\lambda$. 
When the detector is a system with a finite spectrum, the two procedures are no longer equivalent, since $\kappa$ is bounded 
(in the case of a qubit that we are considering, $|\kappa|\le 1/2$). 

We find the probability of postselection (we switch back to manifestly invariant expressions), 
\begin{equation}\label{eq:ppost2}
\P(E_\mathrm{f}) = \frac{1}{2}\left[(1+C^2+S^2 \mathbf{p}\!\cdot\!\mathbf{n})\omega+2 S C(1-\mathbf{p}\!\cdot\!\mathbf{n})\alpha''+
S^2(1-\mathbf{p}\!\cdot\!\mathbf{n})\beta\right] ,
\end{equation}
where $C=\cos{(\lambda/2)}$ and $S=\sin{(\lambda/2)}$. 

The average conditional output is 
\begin{align}
&\langle\tau\rangle= \left[(1+C^2+S^2 \mathbf{p}\!\cdot\!\mathbf{n})+2 S C (1-\mathbf{p}\!\cdot\!\mathbf{n})A''_w+
S^2(1-\mathbf{p}\!\cdot\!\mathbf{n})B_w\right]^{-1}
\nonumber
\\
&\times 
\biggl\{2C^2  \mathbf{q}\!\cdot\!\mathbf{n}+S^2\mathbf{q}\!\cdot\!\mathbf{p}\, (1+\mathbf{p}\!\cdot\!\mathbf{n})
-2SC(\mathbf{q}\!\times\!\mathbf{n})\!\cdot\!\mathbf{p}
\nonumber
\\
&+
2A'_w\left[S^2 \left(\mathbf{q}\!\cdot\!\mathbf{n}-
\mathbf{q}\!\cdot\!\mathbf{p}\, \mathbf{p}\!\cdot\!\mathbf{n}\right)+SC(\mathbf{q}\!\times\!\mathbf{n})\!\cdot\!\mathbf{p}
\right] 
\nonumber
\\
&+
2A''_w\left[SC \left(\mathbf{q}\!\cdot\!\mathbf{n}-
\mathbf{q}\!\cdot\!\mathbf{p}\right)-S^2(\mathbf{q}\!\times\!\mathbf{n})\!\cdot\!\mathbf{p}
\right]
\nonumber
\\
&-
B_w S^2 \mathbf{q}\!\cdot\!\mathbf{p}(1-\mathbf{p}\!\cdot\!\mathbf{n})
\biggr\}.
\label{eq:qqcond}
\end{align}
For $\lambda=\pi$, the results of the preceding subsection are recovered. 

In the regime where we can expand up to first order in $\lambda$, 
\begin{align}
&\langle\tau\rangle\simeq \mathbf{q}\!\cdot\!\mathbf{n} +\frac{\lambda}{2}\left[
(A'_w-1)(\mathbf{q}\!\times\!\mathbf{n})\!\cdot\!\mathbf{p}
-A''_w (
\mathbf{q}\!\cdot\!\mathbf{p}-\mathbf{q}\!\cdot\!\mathbf{n}\mathbf{p}\!\cdot\!\mathbf{n})\right].
\end{align}
Hence, by having $\mathbf{q}$, $\mathbf{n}$, and $\mathbf{p}$ mutually orthogonal, 
the average outcome is proportional to $A'_w-1$; instead, having $\mathbf{q}$ and $\mathbf{n}$ orthogonal 
and $\mathbf{q}$ and $\mathbf{p}$ parallel, the average output is proportional to the imaginary part $A''_w$. 

%
\section{Measurement of a qubit with a general probe.}
This case admits an exact solution, already provided in Ref.~\cite{Duck1989} for a simple case (a probe in a pure Gaussian state interacting instantaneously with the system). 
This solution has been extended to the case of a finite-duration nondemolition measurement with the probe in a mixed Gaussian state \cite{DiLorenzo2008}, the position and value of the maximum output were estimated \cite{DiLorenzo2008}, 
and the solution was also generalized  
to an arbitrary initial state of the probe \cite{DiLorenzo2012a}. Some limiting cases of the exact solution
have also been rediscovered several times \cite{Geszti2010,Wu2011,Zhu2011,Pan2012,Pang2012}. It has been noticed that the exact solution for a spin 1/2 holds also for any operator satisfying $\Hat{A}^2=1$ \cite{Nakamura2012,Susa2012}. 
Furthermore, since $\Hat{\Pi}=(1-\Hat{A})/2$ is a projection operator, the results apply as well to yes-no measurements, which are usually treated separately \cite{Parks2010}. 

In the following we provide compact formulas for this latter case, as a reference for future works on weak measurement 
and to avoid the publication of further duplicated or incremental results. 
The key observation is that \cite{Susa2012} 
\begin{equation}
\Hat{A}^2=1\implies \exp[ix\Hat{A}] = \cos{(x)}+i\sin{(x)} \Hat{A} .
\end{equation}
Thus, for the instantaneous interaction $H=-\lambda\delta(t)\Hat{A}\Hat{P}$, where $\Hat{P}$ is a variable of the detector, the exact joint final state for system and probe is simply
\begin{equation}
{\rho}_{S,P} = \left[ \cos(\lambda\Hat{P}) +i\sin{(\lambda\Hat{P})} \Hat{A}\right] {\rho}_\mathrm{i}\otimes \rho_\mathrm{P}  \left[ \cos(\lambda\Hat{P}) -i\sin{(\lambda\Hat{P})} \Hat{A}\right] .
\end{equation}
Upon postselecting the system in the state $E_\mathrm{f}$, the conditional state of the probe becomes 
\begin{align}
{\rho}_\mathrm{P|f} =& \P(E_\mathrm{f})^{-1} \Tr_S{\left\{(E_\mathrm{f}\otimes\id) {\rho}_{S,P}\right\}}
\nonumber \\
=& \omega \cos(\lambda\Hat{P}) \rho_\mathrm{P} \cos(\lambda\Hat{P})  -i\alpha^* \cos(\lambda\Hat{P}) \rho_\mathrm{P} \sin(\lambda\Hat{P})
+i\alpha \sin(\lambda\Hat{P}) \rho_\mathrm{P} \cos(\lambda\Hat{P})+\beta \sin(\lambda\Hat{P}) \rho_\mathrm{P} \sin(\lambda\Hat{P})
\label{eq:fincondst}
\end{align}
The probability of postselection is 
\begin{equation}\label{eq:probpost}
P(E_\mathrm{f})=\frac{\omega+\beta}{2}
  -\mathrm{Im}\left(\alpha\right) \overline{\sin(2\lambda P)}
+\frac{\omega-\beta}{2} \overline{\cos(2\lambda P)}, 
\end{equation}
where the bar indicates an average over the inital state of the probe.  
After dividing both Eq.~\eqref{eq:fincondst} and Eq.~\eqref{eq:probpost} by $\omega$, 
\begin{align}
{\rho}_\mathrm{P|f} =& 
\frac{ \cos(\lambda\Hat{P}) \rho_\mathrm{P} \cos(\lambda\Hat{P})  -iA_w \cos(\lambda\Hat{P}) \rho_\mathrm{P} \sin(\lambda\Hat{P})
+iA_w^* \sin(\lambda\Hat{P}) \rho_\mathrm{P} \cos(\lambda\Hat{P})+B_w \sin(\lambda\Hat{P}) \rho_\mathrm{P} \sin(\lambda\Hat{P})}
{(1+B_w)/2
  -A''_w \overline{\sin(2\lambda P)}
+(1-B_w) \overline{\cos(2\lambda P)}/2}
\label{eq:fincondst2}
\end{align}
This expression can be used to obtain the average of any observable $\Hat{O}$ of the probe, for any preparation and postselection, and for any coupling strength.

For example, 
the average of a function $f(\Hat{P})$ is 
\begin{align}
\Tr_\mathrm{P}\{f(\Hat{P})\rho_\mathrm{P|f}\}=& \tilde{N}^{ -1}\biggl\{
-A''_w \overline{f(P)\sin{(2\lambda P)}}
\nonumber
\\
&+\frac{1+B_w}{2}\overline{f(P)}
+\frac{1-B_w}{2} \overline{f(P)\cos{(2\lambda P)}}\biggr\},
\label{eq:avfp}
\end{align}
with $\tilde{N}=\P(E_\mathrm{f})/\omega$ the denominator in Eq.~\eqref{eq:fincondst2}. 

\subsection*{Intermission: generalized Wigner functions}
In order to provide the more general result, we need to introduce the generalized Wigner functions. 
We recall that no assumptions are being made on the variable $\Hat{P}$, in contrast to the literature, where it is 
assumed that it has a continuous unbounded spectrum.   
Given an additional operator $\Hat{O}$, with eigenstates $|O\rangle$, we define the generalized Wigner function
\begin{equation}
W_{\Hat{O},\Hat{P}}(O,P) = \int_{D_\mathrm{P}} d\mu_-(p) \langle O|P+p/2\rangle \langle P+p/2|\rho_\mathrm{P}|P-p/2\rangle \langle P-p/2|O\rangle
\end{equation}
where the spectral measure of $\Hat{P}$ is $\mu(P)$, and we defined $\mu_-$ as resulting from a change of variables
 $P=(P_1+P_2)/2$, $p=P_1-P_2$ 
\begin{equation}
d\mu(P_1) d\mu(P_2)=d\mu_+\left(\frac{P_1+P_2}{2}\right) d\mu_-(P_1-P_2) .
\end{equation}
This equation defines at the same time the spectral measure of $P$. We remark that in general, the domain of integration of $p$ depends on 
$P$, unless $\Hat{P}$ has a continuous unbounded spectrum. 
Furthermore, the generalized Wigner functions are real, as the spectrum of $p$ is necessarily symmetric.

Notice that if $\Hat{P}$ is assumed to be a momentum operator, and $\Hat{O}=\Hat{Q}$ to be its conjugated variable, then 
$W_{\Hat{Q},\Hat{P}}$ is but the ordinary Wigner function $W$. Furthermore, in this case $W_{\Hat{Q},\Hat{P}}(Q,P)=W_{\Hat{P},\Hat{Q}}(P,Q)$, but 
in general $W_{\Hat{O},\Hat{P}}(O,P)\neq W_{\Hat{P},\Hat{O}}(P,O)$. 
Notice also that if $[\Hat{O},\Hat{P}]=0$, then $W_{\Hat{O},\Hat{P}}(O,P)=\rho_\mathrm{P}(P,P) \delta(P-O)$. 
There are some fine points arising when the spectrum of either operator is degenerate, which are beyond the scope of the present manuscript. 

The generalized averages are defined as 
\begin{equation}
\overline{F(O,P)}\equiv \int d\mu(O)d\mu_+(P) F(O,P) W_{\Hat{O},\Hat{P}}(O,P)
\end{equation}
with $\mu(O)$ the spectral density of $\Hat{O}$.
In particular, if $F$ is a function of $O$ or $P$ only, then the generalized average reduces to the ordinary one 
\begin{equation}
\overline{F(O)}=\Tr{[ F(\Hat{O}) \rho_\mathrm{P}]} \ , \ \overline{F(P)}=\Tr{[ F(\Hat{P}) \rho_\mathrm{P}]}
\end{equation}
\textbf{End of the intermission}\\

Now we are in the position of providing the conditional average of any observable $\Hat{O}$. 
 The average of a function $g(\Hat{O})$  is 
\begin{align}
\Tr_\mathrm{P}\{g(\Hat{O})\rho_\mathrm{P|f}\}=& \tilde{N}^{ -1}\biggl\{
\frac{1}{2}A'_w\left[\overline{g(O_\lambda)}-\overline{g(O_{-\lambda})}\right]-A''_w \overline{g(O)\sin{(2\lambda P)}}
\nonumber
\\
&+\frac{1+B_w}{2}\frac{\overline{g(O_\lambda)}+\overline{g(O_{-\lambda})}}{2}
+\frac{1-B_w}{2} \overline{g(O)\cos{(2\lambda P)}}\biggr\},
\label{eq:avfo}
\end{align}
where 
\begin{equation}
\Hat{O}_{\pm \lambda} =e^{\pm i\lambda\Hat{P}}\Hat{O}e^{\mp i\lambda\Hat{P}}
\end{equation}
is the operator $\Hat{O}$ in the Heisenberg picture. 

In particular, 
if we assume that $\Hat{P}$ has a continuous unbounded spectrum, so that we may define its conjugate operator 
$\Hat{Q}$, 
the average of a function $g(\Hat{Q})$  is 
\begin{align}
\Tr_\mathrm{P}\{g(\Hat{Q})\rho_\mathrm{P|f}\}=& \tilde{N}^{ -1}\biggl\{
\frac{1}{2}A'_w\left[\overline{g(Q+\lambda)}-\overline{g(Q-\lambda)}\right]-A''_w \overline{g(Q)\sin{(2\lambda P)}}
\nonumber
\\
&+\frac{1+B_w}{2}\frac{\overline{g(Q+\lambda)}+\overline{g(Q-\lambda)}}{2}
+\frac{1-B_w}{2} \overline{g(Q)\cos{(2\lambda P)}}\biggr\},
\label{eq:avfq}
\end{align}
where the overline now indicates an average weighted with the ordinary Wigner function. 

Reference \cite{Pang2012} studies the limit $B_w\gg A_w$ and writes Eqs.~\eqref{eq:probpost}, \eqref{eq:avfp}, and  \eqref{eq:avfq},  in terms of Taylor series, without 
realizing the series could be resummed to give a closed form, as was done in Ref.~\cite{DiLorenzo2012a}. 

Finally, we note that, as the results presented here are valid for any coupling strength, we could have simplified 
the formulas by the rescaling $\Hat{P}\to \lambda \Hat{P}$. However, we avoided doing so, in order for the reader to be able to compare easily 
these general results with those obtained in the weak coupling limit. 
%
\section{Measurement of a general system with a qubit}
We shall now consider the measurement of an observable $\Hat{A}$. 
Such kind of measurement was used by Kocsis \emph{et al.} to determine the Bohmian trajectories of photons \cite{Kocsis2011}. 
The interaction is now taken to be 
$H_\mathrm{int}=-\lambda \hbar\delta(t)\Hat{A} (1-\mathbf{p}\cdot\Hat{\boldsymbol{\tau}})/2$. 
As before, the system is prepared in a state $\rho_\mathrm{i}$ and postselected in $E_\mathrm{f}$. 
The initial state of the qubit probe is 
\begin{equation}
\rho_\mathrm{P} = \frac{1}{2}\left(1+\mathbf{n}\cdot\Hat{\boldsymbol{\tau}}\right).
\end{equation}
\subsection{Approximate results: weak measurement regime}
The weak measurement consists in applying perturbation theory to the joint probability $\P(E_\mathrm{f},\tau)$ and to the 
probability of postselection $\P(E_\mathrm{f})=\sum_\tau \P(E_\mathrm{f},\tau)$, then making a Taylor expansion, if applicable, of 
$\P(E_\mathrm{f},\tau)/\P(E_\mathrm{f})$. 
We expand the time-evolution operator up to first-order terms in $\lambda$. 
Then, 
\begin{align}
\P(E_\mathrm{f},\tau)\simeq& 
\Tr\biggl\{(E_\mathrm{f}\otimes|\mathbf{q}:\tau\rangle\langle\mathbf{q}:\tau|)
\left[1+i\lambda\Hat{A}\frac{1-\mathbf{p}\cdot\Hat{\boldsymbol{\tau}}}{2}\right]
\nonumber
\\
&\qquad\times(\rho_\mathrm{i}\otimes\rho_\mathrm{P})
\left[1-i\lambda\Hat{A}\frac{1-\mathbf{p}\cdot\Hat{\boldsymbol{\tau}}}{2}\right]\biggr\}
\nonumber
\\
\simeq&\ \frac{1}{2}\biggl\{\omega (1+\tau\mathbf{q}\!\cdot\!\mathbf{n})
\nonumber
\\
&\qquad
+i\frac{\lambda}{2}\alpha[1-\mathbf{p}\!\cdot\!\mathbf{n}+\tau(
\mathbf{q}\!\cdot\!\mathbf{n}-\mathbf{q}\!\cdot\!\mathbf{p}+i\mathbf{q}\!\times\!\mathbf{n}\!\cdot\!\mathbf{p})]+c.c.
\nonumber
\\
&\qquad
+\frac{\lambda^2}{2}\beta[1-\mathbf{p}\!\cdot\!\mathbf{n}
+\tau(\mathbf{q}\!\cdot\!\mathbf{p}\, \mathbf{p}\!\cdot\!\mathbf{n}-\mathbf{q}\!\cdot\!\mathbf{p})]
\biggr\},
\end{align}
and 
\begin{align}
\P(E_\mathrm{f})\simeq& \ \omega \left[1-\lambda (1-\mathbf{p}\!\cdot\!\mathbf{n})A''_w+\frac{1}{2}\lambda^2
(1-\mathbf{p}\!\cdot\!\mathbf{n}) B_w\right],
\end{align}
A fundamental consideration is that, while the time-evolution propagator is expanded to first order, the second order
term arising when multiplying $U$ and $U^\dagger$ must be retained, in order for the probability to be positive-definite \cite{DiLorenzo2012j}. 
This prescription is at odds with the naive Taylor expansion learnt from calculus textbook, where the approximation 
applies to functions that do not have to obey any constraint, contrary to a probability distribution. 

The conditional average output of the qubit meter is thus 
\begin{align}
\langle \tau \rangle\simeq \frac{\mathbf{q}\!\cdot\!\mathbf{n}
-\lambda  (\mathbf{q}\!\times\!\mathbf{n})\!\cdot\!\mathbf{p} A'_w
-\lambda(\mathbf{q}\!\cdot\!\mathbf{n}-\mathbf{q}\!\cdot\!\mathbf{p})A''_w-\frac{1}{2}\lambda^2 \mathbf{q}\!\cdot\!\mathbf{p}(1-\mathbf{p}\!\cdot\!\mathbf{n})B_w}
{1-\lambda (1-\mathbf{p}\!\cdot\!\mathbf{n})A''_w+\frac{1}{2}\lambda^2
(1-\mathbf{p}\!\cdot\!\mathbf{n}) B_w}.
\label{eq:genav2}
\end{align}
Notice that in section \ref{sec:qq} the measured variable is $(1-\bm{a}\cdot\Hat{\bm{\sigma}})/2$, 
but the normal weak values $\alpha_{m,n}$ are 
defined in terms of $\bm{a}\cdot\Hat{\bm{\sigma}}$. 
Thus, in order to recover the results of Sec. \ref{sec:qq}, you should put $A_w\to (1-A_w)/2$ and 
$B_w\to (B_w-2A'_w+1)/4$. 

In the linear regime, we recover the results of Ref.~\cite{Wu2009}
\begin{align}
\langle \tau \rangle\simeq \mathbf{q}\!\cdot\!\mathbf{n}-\lambda  (\mathbf{q}\!\times\!\mathbf{n})\!\cdot\!\mathbf{p} A'_w
+\lambda(\mathbf{q}\!\cdot\!\mathbf{p}-\mathbf{q}\!\cdot\!\mathbf{n}\, \mathbf{p}\!\cdot\!\mathbf{n})A''_w.
\label{eq:genav1}
\end{align}
As in the previous section, preparing, interacting, and measuring the qubit meter in orthogonal bases provides $A'_w$, 
while having the measurement and the interaction bases to coincide (or better, not to be mutually unbiased, which suffices) 
provides $A''_w$. 	
\subsection{Exact results}
Equation \eqref{eq:genav2} is an approximate expression, while the analogue relations in the previous 
sections are exact. 
In the following, we shall generalize the approach of Kedem and Vaidman \cite{Kedem2010} in order to find exact expressions for the conditional 
probability of a readout $\tau$ of the qubit probe in the direction $\mathbf{q}$, given that the postselection of the system in the state $E_\mathrm{f}$ was successful. 
Born's rule, applied to the evolved joint state of system and probe, yields 
\begin{align}
\P(E_\mathrm{f},\tau)&=\Tr\left\{\left(E_\mathrm{f} \otimes|\mathbf{q}:\tau\rangle \langle \mathbf{q}:\tau|\right) 
e^{i\lambda \Hat{A} (1-\mathbf{p}\cdot\Hat{\boldsymbol{\tau}})/2}\left(\rho_\mathrm{i}\otimes\rho_\mathrm{P}\right)
e^{-i\lambda \Hat{A} (1-\mathbf{p}\cdot\Hat{\boldsymbol{\tau}})/2}
\right\}
\nonumber
\\
&=\sum_{\tau_1,\tau_2} \langle \mathbf{q}:\tau|\mathbf{p}:\tau_1\rangle 
 \langle \mathbf{p}:\tau_1|\frac{1}{2}\left(1+\mathbf{n}\cdot\Hat{\boldsymbol{\tau}}\right)|\mathbf{p}:\tau_2\rangle 
 \langle \mathbf{p}:\tau_2|\mathbf{q}:\tau\rangle \Gamma_{\tau_1,\tau_2},
\end{align}
where we introduced twice the identity as $\sum_\tau |\mathbf{p}:\tau\rangle\langle\mathbf{p}:\tau|=\id$ and defined 
\begin{equation}
\Gamma_{\tau_1,\tau_2}\equiv \Tr_\mathrm{S}\left\{E_\mathrm{f} e^{i\lambda \Hat{A} (1-\tau_1)/2}
\rho_\mathrm{i} e^{-i\lambda \Hat{A} (1-\tau_2)/2}\right\}. 
\end{equation}
In particular, $\Gamma_{+,+}=\omega$, and, for pure preparation and postselection, $\Gamma_{\tau_1,\tau_2}$ factorizes 
as $\Gamma_{\tau_1,\tau_2}=\gamma_{\tau_1}\gamma^*_{\tau_2}$, with $\gamma$
the propagation amplitude 
$\gamma_{\tau}=\langle \psi_\mathrm{f}| \exp{[i\lambda\Hat{A}(1-\tau)/2]}\;|\psi_\mathrm{i}\rangle$. 
The conditional probability is thus 
\begin{align}
\mathcal{Q}(\tau)
&=\frac{\sum_{\tau_1,\tau_2} \langle \mathbf{q}:\tau|\mathbf{p}:\tau_1\rangle 
 \langle \mathbf{p}:\tau_2|\frac{1}{2}\left(1+\mathbf{n}\cdot\Hat{\boldsymbol{\tau}}\right)|\mathbf{p}:\tau_1\rangle 
 \langle \mathbf{p}:\tau_2|\mathbf{q}:\tau\rangle \Gamma_{\tau_1,\tau_2}}{\sum_\tau \frac{1}{2}(1+\tau\mathbf{n}\cdot\mathbf{p})\Gamma_{\tau,\tau}}.
\end{align}
As elsewhere, one parameter can be eliminated in the conditional probability, for instance dividing numerator and denominator 
by $\omega$
\begin{align}
\mathcal{Q}(\tau)
&=\frac{\sum_{\tau_1,\tau_2} \langle \mathbf{q}:\tau|\mathbf{p}:\tau_1\rangle 
 \langle \mathbf{p}:\tau_2|\frac{1}{2}\left(1+\mathbf{n}\cdot\Hat{\boldsymbol{\tau}}\right)|\mathbf{p}:\tau_1\rangle 
 \langle \mathbf{p}:\tau_2|\mathbf{q}:\tau\rangle C_{\tau_1,\tau_2}}{\sum_\tau \frac{1}{2}(1+\tau\mathbf{n}\cdot\mathbf{p})C_{\tau,\tau}}.
\label{eq:exsolqg}
\end{align}
with $C_{\tau_1,\tau_2}\equiv \Gamma_{\tau_1,\tau_2}/\Gamma_{+,+}$. 
For pure preparation and postselection, $C_{\tau_1,\tau_2}=c_{\tau_1}c^*_{\tau_2}$, 
with $c_+=1$ and $c_{-}=\langle \psi_\mathrm{f}| \exp{[-i\lambda\Hat{A}]}\;|\psi_\mathrm{i}\rangle/\langle \psi_\mathrm{f}|\psi_\mathrm{i}\rangle$. 
The latter term was 
dubbed the modular weak value in Ref.~\cite{Kedem2010}. 
Formally, Eq.~\eqref{eq:exsolqg} is an exact solution. However, it relies on calculating terms like  
$\langle \psi_\mathrm{f}| \exp{[-i\lambda\Hat{A}]}\;|\psi_\mathrm{i}\rangle$. If this term could be evaluated efficiently, then we would have 
an exact solution for any probe, not only the qubit probe being considered here. Generally, this is not the case, as 
the exponential of the operator $\Hat{A}$ is difficult to estimate, if we except some simple cases, as, e.g., the case 
$\Hat{A^2}=\Hat{A}$ considered in Section \ref{sec:qq}, or, more importantly perhaps, the case of $\Hat{A}$ being the 
generator of a transformation. 
%

%

\bibliography{../weakmeasbiblio}
\end{document}